\documentclass[aip,apl,preprint,article,showkeys,prb,citeautoscript]{revtex4} 
\usepackage{graphicx}
\usepackage{bm}
\usepackage{amsmath}
\usepackage{epstopdf}
\usepackage{dcolumn}

\begin{document}

\title{Probing the Quantum States of a Single Atom Transistor at Microwave Frequencies}
\author{Giuseppe Carlo Tettamanzi$^{1}$}
\author{Samuel James Hile$^{1}$}
\author{Matthew Gregory House$^{1}$}
\author{Martin Fuechsle$^{1}$}
\author{Sven Rogge$^{1}$}
\author{Michelle Y. Simmons$^{1}$}

\affiliation{$^{1}$School of Physics and Centre of Excellence for Quantum Computation and Communication Technology, UNSW Australia, Sydney, New South Wales 2052, Australia}

\email{g.tettamanzi@unsw.edu.au}

\date{\today}
\begin{abstract}

The ability to apply~GHz~frequencies to control the quantum state of a single $P$ atom is an essential requirement for the fast gate pulsing needed for qubit control in donor based silicon quantum computation. Here we demonstrate this with nanosecond accuracy in an all epitaxial single atom transistor by applying excitation signals at frequencies up to~$\approx$~13~GHz to heavily phosphorous doped silicon leads. These measurements allow the differentiation between the excited states of the single atom and the density of states in the one dimensional leads. Our pulse spectroscopy experiments confirm the presence of an excited state at an energy~$\approx$~9~meV consistent with the first excited state of a single $P$~donor in silicon. The relaxation rate of this first excited state to ground is estimated to be larger than~2.5~GHz, consistent with theoretical predictions. These results represent a systematic investigation of how an atomically precise single atom transistor device behaves under rf excitations. 
\end{abstract}



\maketitle

\begin{figure*} 
\begin{center}
\includegraphics[width=162mm]{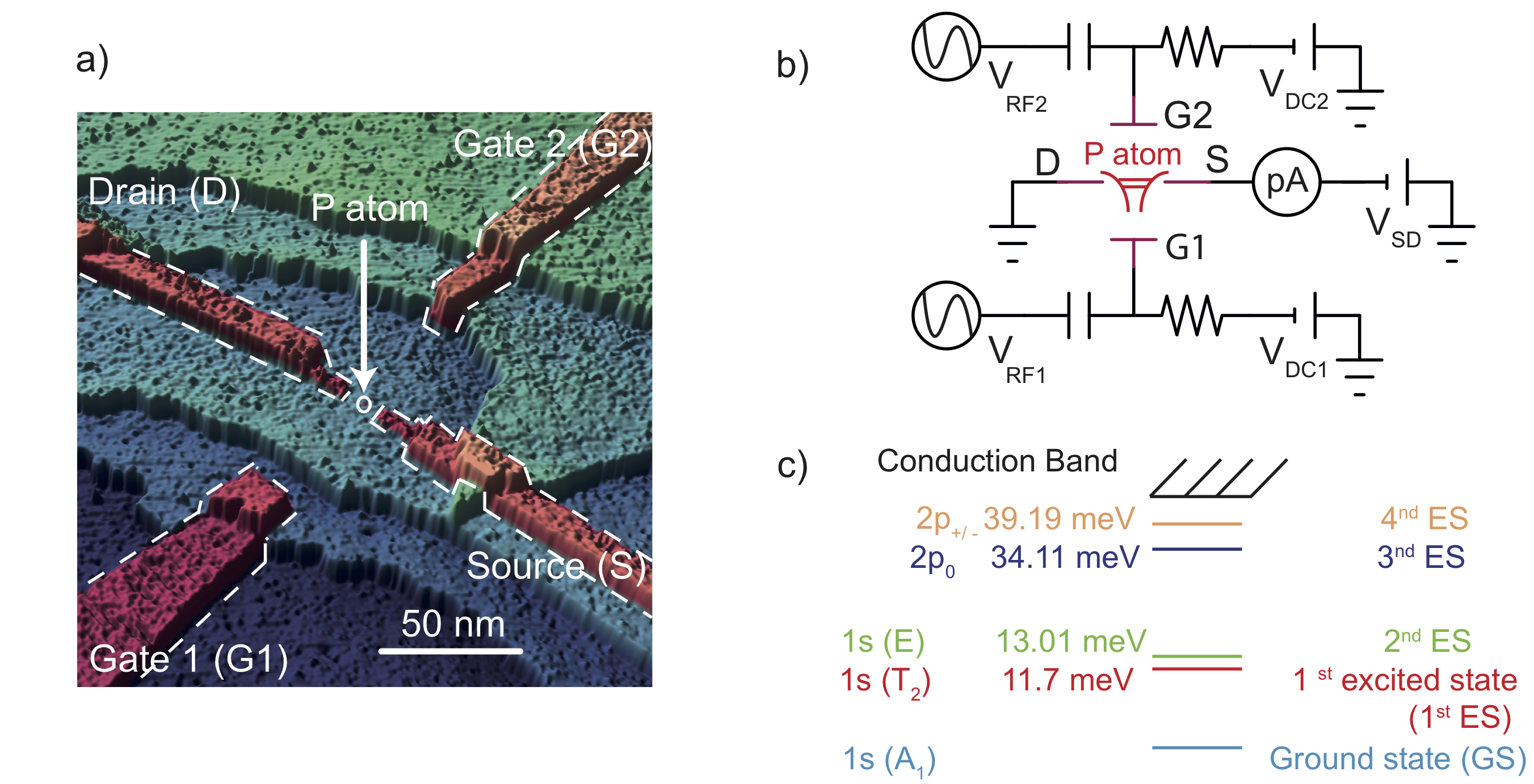}
\caption{~\textbf{High frequency measurements of a single atom transistor}. a)~A scanning tunnelling microscope (STM) image of the device. b)~A schematic of the measurement circuit showing how source-drain leads and two gates lines (G1/G2) are used to control the chemical potential of the donor, where an rf signal $V_{RF1}$/$V_{RF2}$ is added to the conventional dc signal \textit{via} bias-tees. c)~The excitation spectrum of the $D^0$ state of a $P$ donor in Si (in color).~\cite{Fue242,Fue2011,Ram1297} The bulk values for the energy differences between these excited states, ES's, and the ground state (GS-blue) are also shown for the $1^{st}$~ES (red), the $2^{nd}$~ES (green), the $3^{rd}$~ES (dark blue) and the $4^{th}$~ES (orange).~\cite{Fue242,Fue2011,Ram1297} In this picture the lowest three states (1s($A_{1}$); 1s($T_{2}$); 1s(E)) come from the linear combination of the Si valleys due to breaking of valley degeneracy in the Si lattice while the last two (2$p_{0}$ and 2$p_{+/-}$) are orbital-like.~\cite{Ram1297}}
\label{fg:Figuren0}
\end{center}
\end{figure*}

Advances in Si device fabrication technology over the past decade have driven the scale of transistors down to the atomic level. The ultimate limit of this scaling is to fabricate a transistor with just one single dopant atom as the active component of the device and this has been realised using scanning tunnelling microscope (STM) lithography.~\cite{Sch136104} The spin states of individual P donor electrons and nuclei have extremely long coherence times when incorporated into a crystal composed of isotopically purified $^{28}$Si,~\cite{Tyr2012, Ste2012, Muh2014} making them excellent candidates for quantum information processing applications. \cite{Kane133,Loss120,Pla2012} STM lithography offers the potential to scale up such qubits by providing a means to position individual P atoms in a Si lattice, and align them with sub-nanometer precision to monolayer doped control electrodes. This technique has already demonstrated double~\cite{Web4001} and triple~\cite{Wat1830} quantum dot devices, controllable exchange interactions between electrons,~\cite{Web430} and the ability to initialise and read out the spin states of single electrons bound to the donor with extremely high fidelity.~\cite{Wat2015} Most recently, these monolayer doped gates have shown to be immune to background charge fluctuations making them excellent interconnects for silicon based quantum computer.~\cite{Sha233304, Sha236602, Sha3}

Besides the ability to create devices with atomic precision, another requirement for quantum information processing and high-speed logic applications is the ability to control the quantum states of the donor electrons at sub nanosecond timescales. Control signals in the GHz regime are desirable for dispersive readout~\cite{Hou15} and for controlling exchange interactions for non-adiabatic gate operations~\cite{Pet2180}. Indeed, a recently proposed scheme for implementing the surface-code error correction protocol in silicon relies on the ability to propagate signals through such devices with sub-nanosecond timing precision.~\cite{Hill15} Recent impurity-based quantum charge pump devices have been shown to be robust in terms of immunity to pumping errors when operated at GHz frequencies.~\cite{Tet063036,Fuj207} However, to date these experiments have been performed on devices containing random ion implanted impurities.~\cite{Tet063036,Fuj207} STM fabrication capabilities can allow high-precision (~$\lesssim$~nm) positioning of the dopant~\cite{Sch136104}, and when combined with high-speed control of quantum states, it will provide devices for quantum metrology.~\cite{Tet063036,Fuj207}

In this paper we investigate the propagation of high-frequency~signals to the monolayer-doped leads used in atomically precise devices. Previous results have demonstrated the ability to apply radio frequency ($\approx$~$300$~MHz) transmission using dispersive measurements for manipulation of the quantum states.~\cite{Hou15} Here we present a systematic study of the propagation of high frequency signals in atomically precise devices. In this work we demonstrate high frequency capacitive coupling (up to~$\sim$~13~GHz) to the states of a single-atom transistor~\cite{Fue242,Fue2011} fabricated \textit{via} scanning tunnelling microscope lithography, see Fig.~\ref{fg:Figuren0}a), important for the implementation of quantum information processing~\cite{Tyr2012,Ste2012, Muh2014,Kane133,Loss120,Pla2012,Hill15} and quantum metrology.~\cite{Tet063036,Fuj207} We report transient spectroscopy experiments~\cite{Fuj081304, Volk1753} that confirm the existence of the excited state of the P donor located at an energy of~$\approx$~9~meV~$\pm$~1~meV and we extract bounds from 2.5~GHz~to~162~GHz for the relaxation rates from the first excited state to the ground state, $\Gamma_{ES}$, \textit{i.e.} in good agreement with previous experiments~\cite{Zhu093104,HubS211} and theoretical estimations.~\cite{Tah075302} It is important to note that such a large range in the extracted value of~$\Gamma_{ES}$~can be linked to the strong tunnel coupling of the state to source/drain leads in this particular device, making the experiments needed for a more quantitative result infeasible.~\cite{Fuj081304,Volk1753} However, in the long term this coupling can be controlled by the geometry of the tunnel junctions, which can be engineered with sub-nm precision during fabrication.~\cite{Cam2013}

\section*{RESULTS AND DISCUSSION}

In contrast to surface gate-defined quantum dot devices, which typically make use of macroscopic metal electrodes to propagate high-frequency signals, atomic precision devices rely on electrodes formed using highly phosphorus doped silicon ($\sim$~2.5~$\times$~$10^{14}$~$cm^{-2}$) where the phosphorus dopants form a monatomic layer within the Si crystal patterned in the same lithographic step as the single donor atom, see Fig.~\ref{fg:Figuren0}~a). Within the monolayer of dopants the average separation of the donors is~$\lesssim$~1~nm giving rise to a highly disordered two-dimensional electron gas. Disorder scattering in these degenerately doped leads gives rise to a resistance of~$\approx$~hundreds Ohms per square comparable to that found in silicon quantum dots~\cite{Zwa961} but one order of magniture higher than the values observed in conventional transistors.~\cite{Kei7080} However, another very important difference is that the self-capacitance of the atomically thin monolayer wires are negligible with the cross capacitances to the other leads being quite small, estimated to be around the aF~\cite{Fue2011}. As a consequence, very little current ($\approx$~nA) is required to carry a high-frequency voltage signal along these wires if compared to the tens of~nA's necessary for quantum dots.~\cite{Zwa961}

Fig.~\ref{fg:Figuren0} shows in~a)~an STM image of the device and in~b)~a schematic of the measurement circuit used, illustrating how both dc and rf signals can be applied to gate 1 (G1) and to gate 2 (G2) \textit{via} bias-tees. The pink areas in~Fig.~\ref{fg:Figuren0}~a) show the highly $P$ doped monolayer regions (see also methods section) comprising tunnel coupled source/drain (S/D) leads and capacitively coupled gates (G1/G2) surrounding a single phosphorus atom. Several step edges separating the individual atomic planes are clearly visible in the STM image.

\begin{figure*} 
\begin{center}
\includegraphics[width=162mm]{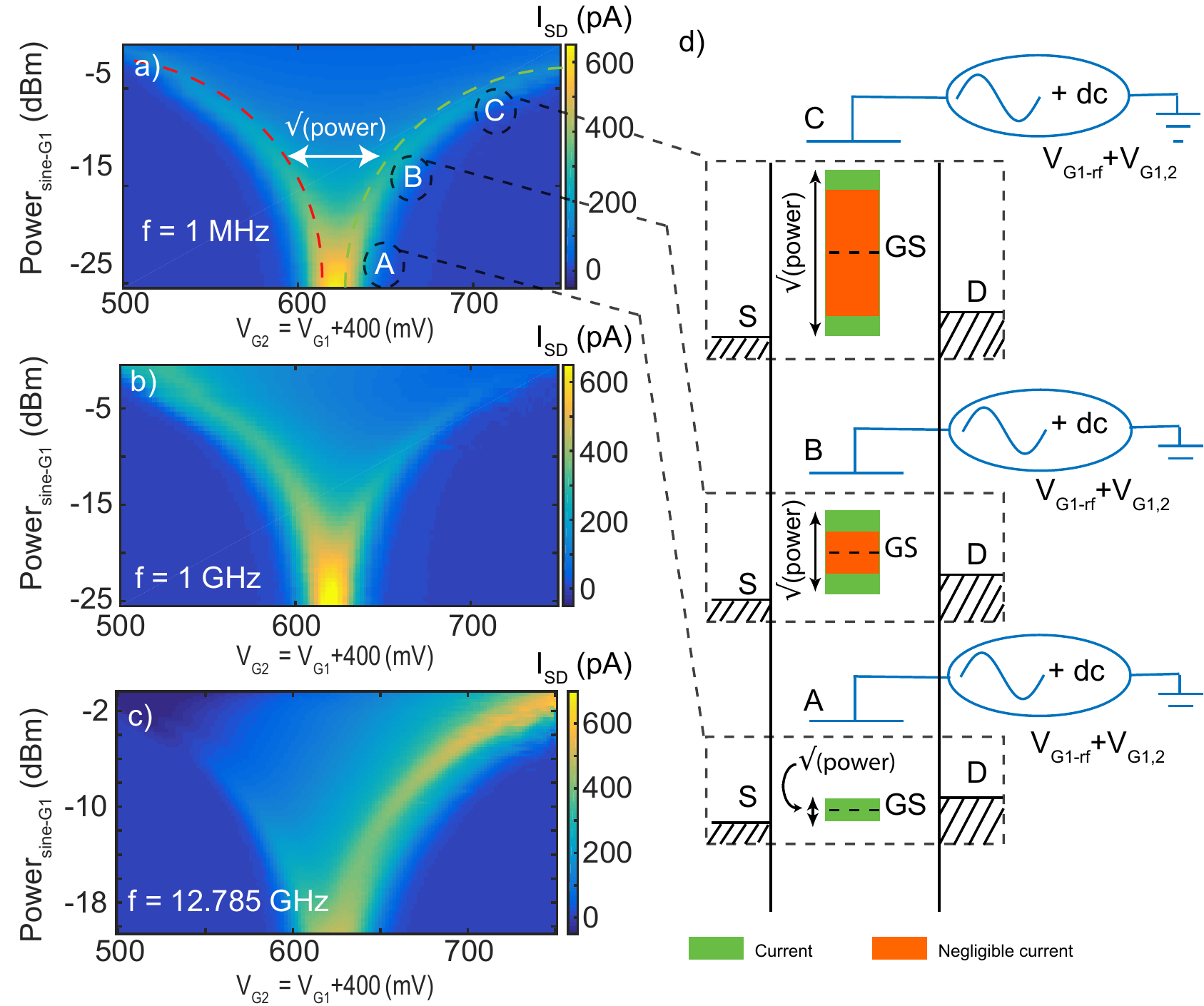}
\caption{~\textbf{High frequency control of the $D^0$ ground state (GS) using the gates (G1, G2).} Square root of the power dependence of the position of the $D^{+}$~to~$D^{0}$ current peak over $\gtrsim$~4 orders of magnitude change in frequency of an rf sine applied to G1 ($V_{G1-{rf}}$), from a)~1~MHz to c)~12.785~GHz where T~=~1.2~K, $V_{SD}$~=~2.6~mV, $V_{Gs}$~=~$V_{G2}$~=~$V_{G1}$~+~400 mV and $V_{Gs_{0}}$ is the position of the GS peak in dc ($V_{G2-0}$~$\approx$~620~mV and $V_{G1-0}$~$\approx$~220~mV). The signal becomes asymmetric above the 1~GHz frequencies, due to frequency dependent cross coupling between the gates and the source/drain leads giving rise to rectification effects. d)~A  schematic describing the doubling of the $D^0$ current peak is shown. As described in the main text, the green/red regions illustrate the positions available for the state when both rf and dc signals are in use.}
\label{fg:Figuren1}
\end{center}
\end{figure*}

To test the frequency response of the monolayer doped gates, the $D^{+}$~to~$D^{0}$ current peak related to current flow through the isolated $P$ atom~\cite{Fue242,Fue2011,Tet063036} can be capacitively addressed by two gates (\textit{i.e.} G1 and G2), allowing an independent rf signal to be added to each of the two gates and the device to be studied in both the dc and the rf domains. The use of rf signals is particularly attractive for these atomic-scale devices as the very narrow leads ($\lesssim$~5~nm) used to address the donor are quasi-1D, make it difficult, by using simple dc bias spectroscopy, to distinguish the signatures in current related to the excited states of the donor from the features related to the density of the states (DOS).~\cite{Fue242,Fue2011,Ryu374,Mot161304} Later we will show how we apply transient current spectroscopy as described in Refs.~\cite{Fuj081304,Volk1753} to clarify some of the transport mechanisms that can arise throughout the excited state spectrum of a single atom transistor.  In Fig.~\ref{fg:Figuren0}~c)~a schematic of the excitation spectrum of the $D^0$ state of a $P$ donor in Si~\cite{Ram1297} is shown highlighting the 1s($A_{1}$), 1s($T_{2}$) and 1s(E) valley states and the 2$p_{0}$ and 2$p_{+/-}$ orbital states of the single donor. In Fig.~\ref{fg:Figuren1} we observe the evolution of the current peak related to the ground state (GS) of the $D^0$ state as a function of the power of the sinusoidal rf signal added to the dc voltage of gate 1. The possibility of capacitively addressing this $D^0$ GS is confirmed for high frequencies up to~$\approx$~13~GHz, where, as expected, when an rf signal with sufficient power is in use, the position of the $D^{0}$ current peak splits in two, with the splitting being proportional to the square root of the power of the provided excitation. This doubling of the current peak is observed for more than 4 orders of magnitude change in the frequency (\textit{i.e.} from~1~MHz to~$\sim$~13~GHz). In Fig.~\ref{fg:Figuren1}d) we show a schematic describing how the doubling appears at different power and the underlying mechanisms causing it. When the rf signal is applied to one of the two gates (G1 or G2), during each rf cycle, the GS can occupy a range of positions represented by the green/red regions in the schematic of Fig.~\ref{fg:Figuren1}d) where the green and red regions simply refer to the voltage change rate at which the donor GS crosses the bias windows and depends on the timing of the sine wave (green~=~low rate of change of the sine; red~=~high rate of change of the sine). 

To clarify, at any point in time of the sine period the current is proportional to the portion of integrated time that the states spend within the bias windows. Hence, if the variation in time of the position of the state is minimal (\textit{i.e}.~$\frac{d[sine(\omega t)]}{dt}|_{\omega t~=\pm 90^o}~\approx~0$), as in the green regions in Fig.~\ref{fg:Figuren1}d), it is possible for electrons to tunnel resonantly between the source and the drain \textit{via} the state and it is possible to observe a current. However, if this variation in time is maximum (\textit{i.e.}~$ \frac{d[sin(\omega t)]}{dt}|_{\omega t~=0^o}$~$\gg$~1), as in the red regions, only negligible current can be observed.

\begin{figure} 
\begin{center}
\includegraphics[width=81mm]{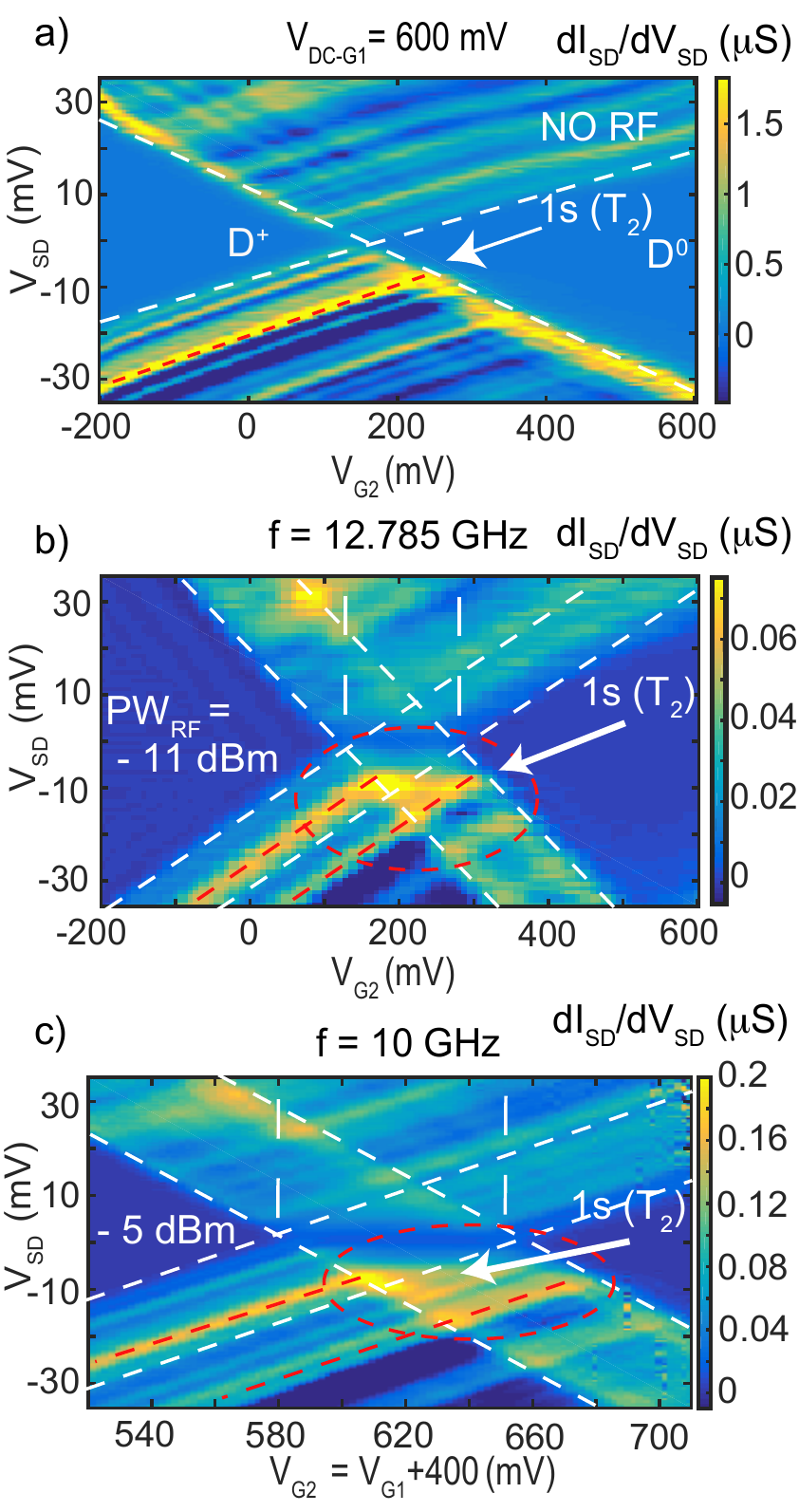}
\caption{~\textbf{Excited state spectrum at high rf frequencies}. a)~A dc gate stability diagram and b)~Power dependence of the excited state spectrum when an rf excitation at $\nu$~=~12.785~GHz is applied to gate 1. c)~Same measurement as in b) but with an rf excitation at $\nu$~=~10~GHz on gate 1 and both the gates addressed in dc as in~Fig.~\ref{fg:Figuren1}. The red ellipses and the red dashed lines in all sections of the figure outline the $1^{st}$~excited state located at 10~meV~$\pm$~2~meV.}
 \label{fg:Figuren3}
\end{center}
\end{figure}

In Fig.~\ref{fg:Figuren3} we now turn to the impact of the rf on the response on the excited states of the donor atom. Fig.~\ref{fg:Figuren3}~a) shows the excited state spectrum at the $D^{+}$~to~$D^{0}$ transition with no rf signal applied also consistent with previous measurements of this device.~\cite{Fue242,Fue2011,Ryu374} In this figure the dc charge stability diagram focusses on the position of the first excited state, 1s~($T_{2}$). Since we have the availability of two gates and the device is highly symmetric, the donor states can be capacitively addressed with both G1 and G2. As a consequence we can address the states in two different regimes either when G1 and G2 are tied together and varied simultaneously or when G1 is fixed and G2 is varied. By addressing the donor in those two different regimes we observe the same spectrum as in the original measurements~\cite{Fue242,Fue2011} where, as expected,~\cite{Rah165314} the positions of each level are insensitive to the changes in electric field related to the different measurement configurations, see Fig.~\ref{fg:Figuren3}a) and Fig.~\ref{fg:Figuren4}c). Obtaining the same results with different regimes of addressing the states is important as it demonstrates the reproducibility under different electric field conditions.

In Fig.~\ref{fg:Figuren3}~b) we now present the same spectrum but with an rf excitation of $\nu$~=~12.785 GHz applied to G1. We observe the same doubling of current signature as observed in Fig.~\ref{fg:Figuren1}c), but now for the excited state spectrum, \textit{e.g.}~1s($T_{2}$). This first excited state is located at 10~meV~$\pm$~2~meV, consistent with the previously measured bulk value for the first excited state of a single P donor in bulk silicon~\cite{Fue242,Fue2011,Ram1297} as shown in Fig.~\ref{fg:Figuren0}c). Likewise in Fig.~\ref{fg:Figuren3}~c) we observe the same effect but now using an rf excitation at $\nu$~=~10 GHz applied to G1 and with both gates addressed in dc, as in Fig.~\ref{fg:Figuren1}. These results are similar to those discussed in Fig.~\ref{fg:Figuren1}d), however this time the capacitive coupling is demonstrated for the 1s($T_{2}$) level of the excited state spectrum and shows robustness to the electric field across the donor.

\begin{figure} 
\begin{center}
\includegraphics[width=81mm]{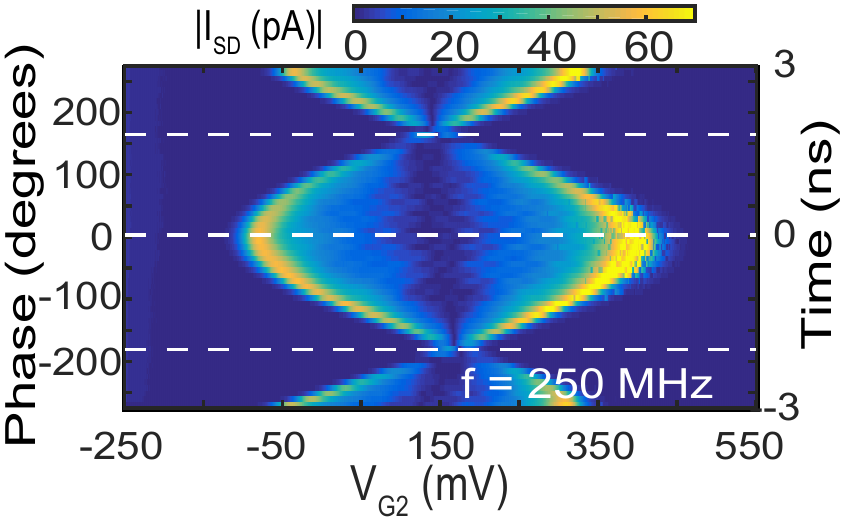}
\caption{~\textbf{Nanosecond synchronisation between G1 and G2}. A dc voltage is applied to G2 while a rf sine excitation of amplitude~$\approx$~-~8~dBm is provided to \textit{both} G1 and G2. $V_{dc-G1}$ and $V_{SD}$~are kept fixed at 600~mV and 0~mV, respectively. As expected the maximum splitting of the peak is observed when the two rf signals are in-phase (0 degrees or 0~ns) while the minimum splitting is observed when they are out-of-phase ($\approx$~$\pm$~180 degrees or~$\pm$2~ns).}
 \label{fg:Figuren3bis}
\end{center}
\end{figure}

Quantum information and quantum metrology applications require precise and independent rf control of different gates,~\cite{Loss120,Hill15,Fuj207} as many quantum logic gate operations include fast manipulation of electron states. These operations require the \textit{absolute} synchronisation in phase (time) between the rf signal individually applied to different gates and of their relative coupling to quantum states. To ascertain if this is possible in our device, in Fig.~\ref{fg:Figuren3bis} we present results obtained from applying sinusoidal rf excitations of 250\,MHz to the bias-tees of both G1 and G2. Here, the provided rf excitations are of equal amplitude but there is a varying difference in the absolute phase between the two signals. Hence, Fig.~\ref{fg:Figuren3bis} ultimately allows to quantify the level of synchronisation in time between the capacitive coupling between G1 and the GS and the one between G2 and the GS. The result of these measurements confirms that, within the limit of precision of the source (~$\approx$~10~ps, see methods section), a very similar capacitive coupling between each gate and the donor state~\cite{Fue2011} is in place and is preserved in the rf regime. These results show that, by precision STM patterning, it is possible to have control of the device symmetry and, as a result, to observe accurate nanosecond synchronisation between different gates up to 0.25~GHz frequencies. The results presented so far are of relevance for the field of quantum computations as they demonstrate the control of energy states at $f$~$\gtrsim$~10~GHz, \textit{i.e.} the high frequencies required for several quantum computer proposals which, require synchronous sub-ns pulses to be applied to quantum states~\cite{Loss120,Hill15}. Precision transistors can also be used for single-electron transfer applications, such as the ones necessary for quantum metrology,~\cite{Tet063036,Fuj207} where independent and precise control in time of more than one gate is needed.

\begin{figure} 
\begin{center}
\includegraphics[width=162mm]{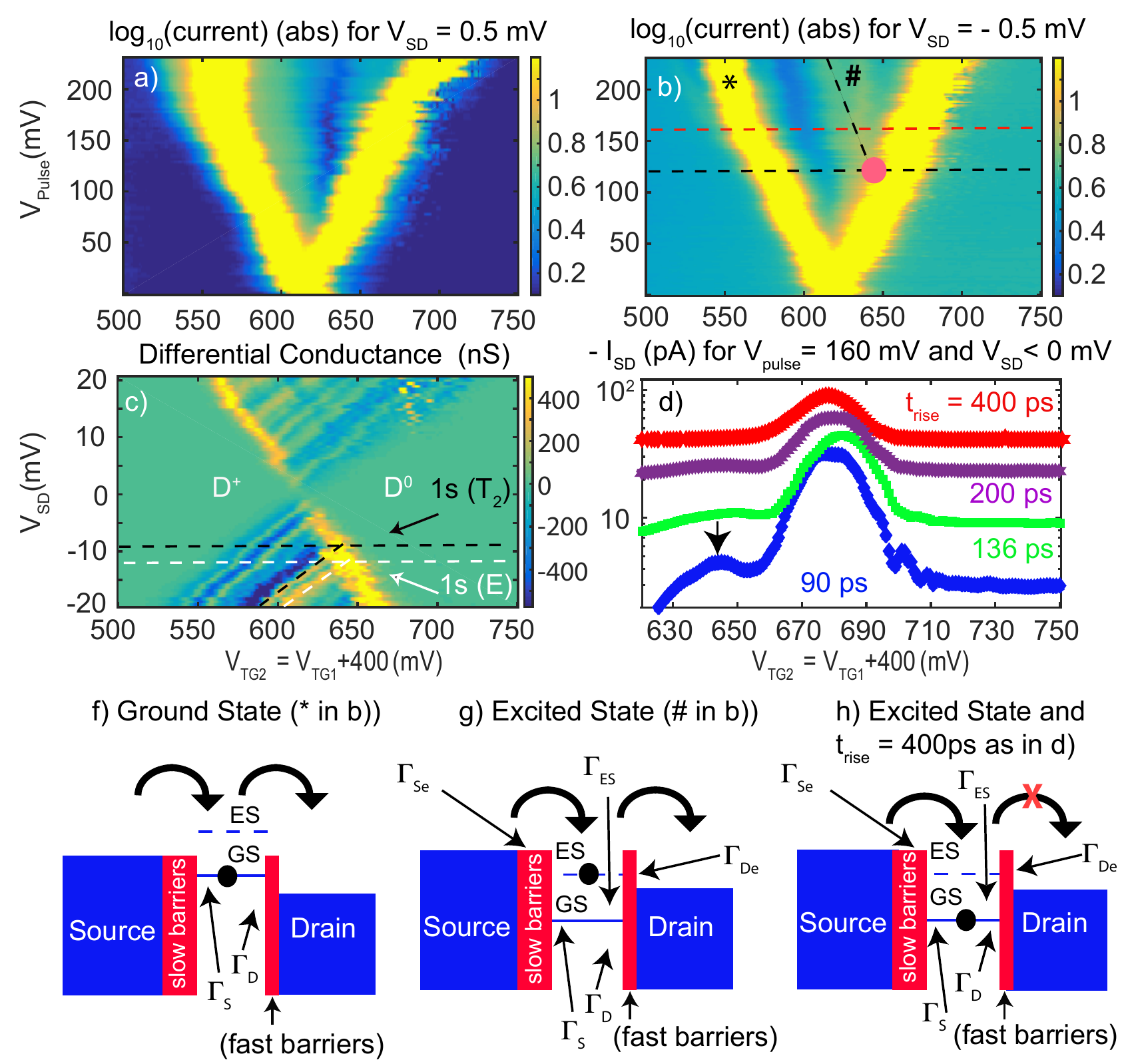}
\caption{~~\textbf{Estimation of the 1s($T_{2}$) excited relaxation rate}. The observed V-shapes represent the doubling of the ground state peak when square pulses with  $\nu$~=~50~MHz, rise times of 90 ps and a duty cycle of 50~$\%$ are applied both for a)~$>$~0 and for b)~$<$~0 bias voltages. c)~Close up of the stability diagram around the $D^{0}$ peak. The black arrow indicates the position of the excited state (1s~($T_{2}$)) responsible for the extra current signature observed in b) outlined with the $\#$ symbol. The white arrow indicates the position of the $2^{nd}$ excited state (1s~(E)). d)~Evolution of the excited state feature for different values of the rise time, $t_{rise}$. The curves represent the logarithm of the measured current which is originally negative and have a small offset added for clarity. f), g) and h)~show a schematic of the pulsing behaviour observed in b) and d) and discussed in the main text.}
 \label{fg:Figuren4}
\end{center}
\end{figure}

In the next section we shall show how, using excited state spectroscopy at~$\nu$~=~50~MHz,~\cite{Fuj081304, Volk1753} we can distinguish the electron excited states spectrum of the donor from the 1D confinement-related DOS of the quasi one dimensional leads.~\cite{Fue242, Fue2011, Ryu374, Mot161304} As in the previous experiments, when we apply a square wave signal to one of gates addressing the state, we observe a characteristic V-shape of the current as a function of increasing pulse voltage (see Fig.~\ref{fg:Figuren4} where $V_{pulse}$ represents the voltage amplitude provided to the bias-tee). The V-shape of the current represents the doubling of the ground state peak when square pulses are applied to G1 and is observed both for positive (Fig.~\ref{fg:Figuren4}a) and for negative (Fig.~\ref{fg:Figuren4}b) source bias voltages. This process is schematically described in Fig.~\ref{fg:Figuren4}f) for negative biases. In Fig.~\ref{fg:Figuren4} a) and b) the left branch shows the current where the ground state is pulsed from far above the bias window, while the right branch represents the dc ground state signature, which is shifted by the introduction of the pulse. There is an additional feature, labelled "$\#$", observed when a negative bias is applied to the source, as in Fig.~\ref{fg:Figuren4}~b), which we attribute to the $1^{st}$ excited state of the donor electron as explained in the next section. It is worthwhile to remember that the DOS in the one dimensional leads cannot be associated with this additional feature because the DOS signature is not $V_{Pulse}$-dependent but only S/D bias-dependent.~\cite{Fue242, Fue2011,Mot161304} Hence, in these experiments we can address both the excited and ground state spectrum \textit{at low bias} such that pulse spectroscopy allows to distinguish transport \textit{via} the excited state and the DOS in the leads in a way not possible \textit{via} dc spectroscopy.~\cite{Fue242, Fue2011} The Coulomb diamonds and the doubling observed in Fig.~\ref{fg:Figuren4} a) and b) allow a direct conversion between gate voltage and energy. From the position of the red dot in Fig.~\ref{fg:Figuren4}b) at $V_{Pulse}$~=~120~mV~$\pm$~10~mV and using~0.075 for the final correction factor of the applied power (see methods section), we can determine an excited state energy of~9~meV~$\pm$~1~meV.

This pulsed-estimated value for the excited state energy 1s($T_{2}$) lies close to the one extracted from the dc data in Fig.~\ref{fg:Figuren4}c) $\approx$~10~meV~$\pm$~2~meV (black arrow and black dashed lines), see also red ellipses and red dashed lines in Fig.~\ref{fg:Figuren3}b) and Fig.~\ref{fg:Figuren3}c). The position of the other visible peak for the excited state (1s(E), white arrow and white dashed line around 13.5 meV), is also very close to the expected bulk values, \textit{i.e.} 11.7~meV and from previous estimations made of this device, in the dc mode.~\cite{Fue242, Fue2011}

It is important to understand why the overall dc excited state spectrum is more visible for negative bias (\textit{e.g.}~see~Fig.~\ref{fg:Figuren4}~c)) which indicates that the transparencies of the source/drain to $1^{st}$ excited state barriers ($\Gamma_{Se}$/$\Gamma_{De}$) are asymmetric, with the latter being more transparent.~\cite{Fue2011,Lan136602} The asymmetry in the tunnel barriers (\textit{i.e.} $\Gamma_{Se}$/$\Gamma_{De}$~$\ll$~1) can be better understood by looking at Fig.~\ref{fg:Figuren4}g) where the negative bias regime is schematically illustrated. This figure shows that if the electrons moving from source to drain \textit{via} the $1^{st}$ excited state encounter first a slow barrier, $\Gamma_{Se}$, and then a fast one, $\Gamma_{De}$, then the dc excited state signature will be more visible compared to the opposite case of positive bias where an electron will first encounter a fast barrier and then a slow one. In the later case electrons are most likely to relax to the ground state before tunnelling through the slow barrier and the excited state signature will be less visible. Furthermore, as the same asymmetry applies for pulsing experiments and, at negative bias and for sufficiently slow relaxation from the 1s ($T_2$) excited state to the ground state, $\Gamma_{ES}$, if compared with $\Gamma_{De}$, we see the excited state line once the pulse amplitude exceeds the ES energy (black dashed line, the pale red dot and the~$\#$~symbol in Fig.~\ref{fg:Figuren4}b)). This because the asymmetry will allows a better visibility of the excited state 1s~($T_{2}$) but this time at low bias and without the presence of the DOS signature complicating the picture. It can be easily seen~\cite{Fue2011} that the same asymmetry observed in the tunnel barriers $\Gamma_{Se}$/$\Gamma_{De}$ is also true for $\Gamma_{S}$/$\Gamma_{D}$ with $\Gamma_{S}$ and $\Gamma_{D}$ being the source to ground state and the drain to ground state barriers, respectively, where the following two inequalities can also be obtained:~\cite{Fuj081304,Volk1753,Tah075302}

\begin{equation}
\label{eq:1}
\Gamma_{Se}~\gtrsim~\Gamma_{S}  
\end{equation}

\begin{equation}
\label{eq:2}
\Gamma_{De}~\gtrsim~\Gamma_{D}
\end{equation}

where these two inequalities are due to the typical larger spatial extent of the excited state wave functions compared to the ground state ones. 
\footnotetext[1]{The estimation of these rates comes from the assumption that, in this device, the transport at low temperatures ($\approx$~100~mK) is in the life time broadening regime which allow to extract a first set of indicative values for the values of $\Gamma_{S}$~$\approx$~150~MHz and $\Gamma_{D}$~$\approx$~164.5~GHz, see also Ref~\cite{Fue2011}.}

The values of the two barriers $\Gamma_{S}$ and $\Gamma_{D}$ have been already quantified \textit{via} a simple modelling\footnotemark[1] to be~$\Gamma_{S}$~$\cong$~150 MHz and~$\Gamma_{D}$~$\cong$~164.5 GHz confirming the expected asymmetry of the barriers ($\Gamma_{S}$/$\Gamma_{D}$~$\approx$~$10^{-3}$), not unusual for these systems.~\cite{Lan136602} Since Eq.~\ref{eq:2} is true, it follows that~$\Gamma_{ES}$~$<<$~$\Gamma_{D}$. As a consequence we can obtain bounds for $\Gamma_{ES}$ from the following points: 

\begin{itemize}
  \item~The rise time from 10~$\%$ to 90~$\%$ of the maximum amplitude~\cite{Fuj081304, Volk1753} of the used pulsed signal is 90~ps (11~GHz), hence the pulse brings the excited state in resonance within this~11~GHz range of frequencies. This gives us the information that~$\Gamma_{S}$ is~$<$~11~GHz in agreement with what already discussed. 
  \item~The amplitude of the excited state signal in Fig.~\ref{fg:Figuren4}b) is $\approx$~4~pA, hence it is possible to estimate that~$\approx$~50~$\%$ of the electrons are loaded \textit{via} the excited state during each individual pulse. Also, the edge of the square pulse~\cite{Fuj081304, Volk1753} can never be sharp as in an ideal case, hence this indicates that $\Gamma_{Se}$ cannot be much faster than $\Gamma_{S}$, in agreement with Equ.~\ref{eq:1}.
  \item~If a positive bias is applied to the device, as in Fig.~\ref{fg:Figuren4}a), the electrons first encounter the fast barrier and then the slow one, $\Gamma_{Se}$. As no extra signal can be observed for this regime, this indicates that the electrons are always relaxing to the ground state before being able to tunnel to the source, leading to the conclusion that $\Gamma_{Se}$~$\ll$~$\Gamma_{ES}$, in agreement with recent theoretical estimations.~\cite{Tah075302}
\end{itemize}

The set of observations just discussed together with and Eq.~\ref{eq:1} and Eq.~\ref{eq:2}  allow us to determine approximative bounds for $\Gamma_{ES}$ as in the following inequalities:
 $\Gamma_{D}$~$\gg$~$\Gamma_{ES}$~$\gg$~$\Gamma_{S}$, hence 164.5~GHz~$\gg$~$\Gamma_{ES}$~$\gg$~150~MHz. To test this hypothesis further, in Fig.~\ref{fg:Figuren4}d) traces are taken for a fixed $V_{Pulsed}$~=~160~mV (as in the red dashed line in Fig.~\ref{fg:Figuren4}b)) across the excited state signal. Here we can see by adding different low pass filters to the pulse line (at room temperature and one at the time) we can change the pulse rise time and observe if the extra signal related to the excited state can be attenuated. In fact, by increasing the rise time of the pulse to 400 ps (\textit{i.e.} by using a~2.5~GHz low pass filter), the extra signal can be completely suppressed. As shown schematically in Fig.~\ref{fg:Figuren4}g) and in ref.,~\cite{Fuj081304, Volk1753} the fast rise time of the pulse is a fundamental requirement for the observation of the resonant tunnelling \textit{via} the excited state. If the rise of the pulse edge is too slow compared to~$\Gamma_{ES}$ and $\Gamma_{S}$, then the electrons cannot resonantly tunnel \textit{via} the excited state but instead have always an higher chance to first tunnel to the ground state cancelling the possibility of observing the extra current signature.
 

Here we argue that the use of the filters and the reduction of the rise time to 400~ps ultimately favours tunnelling \textit{via} the ground state rather than \textit{via} the excited state. This allows us to give a better estimation of the value for~$\Gamma_{S}$~$\approx$~2.5~GHz, since only when the rise time and $\Gamma_{S}$ have similar values can the extra signal related to the resonant tunnelling \textit{via} the excited state be suppressed. Note that this value of $\Gamma_{S}$ is higher than the value of~$\Gamma_{S}$~extracted from dc transport,\footnotemark[1] but still realistic. This correction on the estimation of~$\Gamma_{S}$, leads also to a slightly improved estimation of~$\Gamma_{D}$~$\cong$~162~GHz,\footnotemark[1] while still confirming the asymmetry between the two barrier rates.

The use of filters described above and schematically drawn in Fig.~\ref{fg:Figuren4}h) can provide a \textit{rough} estimate for $\Gamma_{S}$, since it is not easy to determine the final influence that the filter has to the shape of the pulse,~\cite{Fuj081304, Volk1753} however it gives a better indication for the bounds of~$\Gamma_{ES}$. In fact, this discussion suggests that a better defined range for the value of~$\Gamma_{ES}$ is:~162~GHz~$\gg$~$\Gamma_{ES}$~$\gg$~2.5~GHz, which is compatible with theoretical predictions and with experimental observations.~\cite{Tah075302,Zhu093104,HubS211} We have shown how to extract limits for the value of the relaxation rate of the first excited state of an isolated P donor. As traditionally these quantities are difficult to measure~\cite{Fuj081304,Volk1753,Zhu093104,HubS211} or estimate theoretically~\cite{Tah075302,Riw235401} this is a relevant result for the fields of Si quantum information and Si quantum metrology. In these planar doped devices the barriers $\Gamma_{S}$ and $\Gamma_{Se}$ are tuneable only during fabrication allowing us to control the tunnel rates by an order of magnitude with precision lithography using current techniques, with future experiments aimed at improving this further.~\cite{Cam2013} This non tunability of the barriers during experiments represents an ultimate limit to the pulsing frequency that can be used. Hence, the higher pulse frequency, of the same order of magnitude as the relaxation rates~$\approx$~10~GHz, necessary to obtain a quantitative value~\cite{Fuj081304, Volk1753} of $\Gamma_{ES}$ as in Refs.~\cite{Zhu093104,HubS211,Tah075302} is not accessible. However, the regime explored in these experiments demonstrates the potential of the fast pulsing technique with all epitaxial monolayer doped gates. The discussion contained in this last section also explains why no excited state substructure can be observed in this Fig.~\ref{fg:Figuren1}, as the use of a sinusoidal excitation doesn't provide the appropriate conditions (as in Fig.~\ref{fg:Figuren4}g)) for the electrons to resonantly tunnel \textit{via} excited states when the S/D bias is small.

\section*{CONCLUSIONS}

In conclusion, in this work we demonstrated fast rf control of the excited state spectrum of a $P$ atom in a single atom transistor using all epitaxial monolayer doped gates. This control was performed at GHz speed and with nanosecond synchronisation needed to execute quantum gate operations in several silicon based quantum computer proposals. Pulsed spectroscopy measurements with selective transport \textit{via} excited states allowed us to differentiate between the excited states of the single atom and the density of states in the one dimensional leads in a manner not possible \textit{via} dc spectroscopy. From these measurements we demonstrated a possible range of values for the relaxation times from the first excited state to the ground state. Such excited state relaxation rate information will help in the assessment on how realistic is the use of the silicon quantum valley-orbital degree of freedom for quantum logic and quantum metrology applications.~\cite{Riw235401,Tet046803,Cul126804} This work shows that with precision single atom fabrication technologies with epitaxial monolayer doped gates we can apply voltages up to GHz frequencies to control the spin states of the qubits. With the recent demonstration of the suppression of charge noise in these systems~\cite{Sha233304, Sha236602, Sha3} this bodes well for precision donor based qubits in silicon.

\section*{ACKNOWLEDGMENTS}

G. C. Tettamanzi acknowledges financial support from the ARC-Discovery Early Career Research Award (ARC-DECRA) scheme, project title : ÕSingle Atom Based Quantum MetrologyÕ and ID: DE120100702 for the development of the setup used in these experiments. M.Y.S. acknowledge a Laureate Fellowship (FL130100171).

\section*{METHODS AND EXPERIMENTAL}

\textbf{Fabrication of the Single Atom Transistor Device.} The device is fabricated on a low-doped (1-10\,$\Omega$\,cm) silicon wafer prepared with a Si(100)~2x1~surface reconstruction using a flash anneal to 1150\,$^{\circ}$C, before it is passivated by atomic hydrogen. Controlled voltage and current pulses on the STM tip locally desorb this hydrogen layer to define the device features with atomic precision, leaving behind chemically active Si unpaired bonds. PH$_3$ gas introduced into the chamber binds to the surface in the regions where the hydrogen was desorbed. An anneal to 350\,$^{\circ}$C causes the P atoms to incorporate into the top layer of the Si crystal. The P doped features are then encapsulated by low temperature ($\lesssim$~250~$^o$C) solid source Si molecular beam epitaxy. The all epitaxial doped leads are electrically contacted by first using reactive ion etching to etch holes in the encapsulation down to the doped layer, then the holes are filled by evaporation of Al to make ohmic contact to the P doped layer. The P doped leads in this device are~$\sim$~1000~nm long and widen between~$\approx$~5~nm in the central part of the device to 800 nm in the contact region, with in an estimated 36 $k\Omega$ of two terminal resistance along the length of the leads.~\cite{Fue2011} 

\textbf{Low Temperature and rf Measurements.} The device is mounted on a cold finger of $^{4}$He pot of an Oxford Variable Temperature Insert (VTI) operated at~1.2~K. A low noise battery operated measurement setup was used to measure the source/drain current and to apply the dc voltages. To apply the sinusoidal rf input to the gates \textit{via} the bias-tees an Agilent E8257C source (operating up to 40 GHz) and a two channels Agilent 81180A source were used. The inter-channel time skew control of the Agilent 81180A source goes from -3 ns to +3 ns with 10-ps precision and determines the best possible control in time/phase between the two different rf signals (10-ps which is equivalent to 0.9 degrees for the used~$\nu$~= 250 MHz of our experiments). rf signals can be transferred to the bias-tees \textit{via} high performance coax rf lines. 

These lines have silver-plated copper-nickel inner conductor and copper-nickel outer conductor (\textit{i.e.} attenuation ranging between the sub dBm/m to the few dBm/m at 20 GHz). SK coaxial rf connectors are used in all these rf lines and 6 dBm attenuators are placed as close as possible to the bonding pads ($\lesssim$~1 cm). The bias-tees are built with typical resistance and capacitance values of R = 1 M$\Omega$ and C = 1 nF, respectively. The used values for R and for C leads to characteristic RC times of around the few ms and high pass filter cutting frequencies of around a 0.1 KHz. These bias-tees have also been tested independently with a Keysight N9918A FieldFox handheld microwave analyser and have shown to operate with no resonances and with the expected linear increases of the losses up to the 26.5 GHz (\textit{i.e.} the limit of our analyser).

Furthermore, a correction factor~$\cong$~0.75 estimated \textit{via} the~$\frac{\Delta V_{pulsed}}{\Delta V_{Gs}}$~$\approx$~$\frac{150}{200}$ of the V-shape in Fig.~\ref{fg:Figuren4}a) and in Fig.~\ref{fg:Figuren4}b) is used to take into account the attenuation of the signal at the bias-tee level (for~$\nu$~=~50~MHz) while the gate lever arm has been already estimated to~$\approx$~0.1~\cite{Fue2011} making the final correction factor of the applied power equal to 0.075. Indeed, experiments as shown in Fig.~\ref{fg:Figuren1} have been possible up to 20~$\lesssim$~GHz, knowing that neither the rf source or attenuation in the rf lines are a limitation to these experiments for~$\nu$~up to 40 GHz and the bias-tees attenuation is not a limitation to these experiments for~$\nu$~up to 26.5 GHz. The limitation on the maximum frequency of operation of our device is most likely due to imperfect 50~$\Omega$ matching at the interface between the Al/Si bonding wire and the bonding pad of the device (used to connect the device to the external setup). The pulsing experiments have been performed with an HP 8131 and with an Agilent 81180A
AWG in combination with a fast switching optical isolator from Delft university (http://qtwork.tudelft.nl/ schouten/ivvi/doc-mod/docs5d.htm). Overall, the 10~$\%$~to 90~$\%$~rising time was estimated with a fast oscilloscope to be~$\cong$~90~ps for the pulses used in Fig.~\ref{fg:Figuren4}, hence we believe that the AWGÕ is not a limiting factor to the excited state spectroscopy experiments.

\section*{ADDENDUM}

In our paper we provided an upper bound for the relaxation from the 1s ($T_2$) excited state to the ground state, $\Gamma_{ES}$. This was determined from our estimation that the drain to ground state fast tunnel barrier, $\Gamma_{D}$ was equal to 162~GHz. We have now realised that in Figure 5g) of the paper, where the transport mechanisms that allowed the observation of the excited state signature are schematically explained, we neglected to consider the process where an electron tunnels from drain to the ground state, see green arrow in Figure~\ref{fg:Fig} below. This neglected process can lead to blockading of the loading of the excited state, which no longer contributes to a net current through the Single Atom Transistor. The fact that we nevertheless observe the excited state resonance in the experiment suggests that the rate for this loading (i.e. $\Gamma_D$) is not greater than the source to excited state ($\Gamma_{Se}$) as previously estimated and is much smaller than 162~GHz. It does however indicate that these two rates are of the same order of magnitude, and the competition between the two processes contributes to the observed current (4~pA). This interpretation is a minor adjustment to that given in the paper, only requiring $\Gamma_{Se}$ and $\Gamma_D$ being of the same order of magnitude while being much smaller than $\Gamma_{ES}$. In this new interpretation of the data the estimation of the source to ground state slow tunnel barrier, $\Gamma_{S}$, being much smaller of $\Gamma_{D}$ is confirmed. The same applies for the lower bound for $\Gamma_{ES}$ obtained in the paper and discussed in the abstract. However, it is no longer possible to estimate an higher bound for the value of the $\Gamma_{ES}$ which was previously stated as 162 GHz. 

\begin{figure*} 
\begin{center}
\includegraphics[width=162mm]{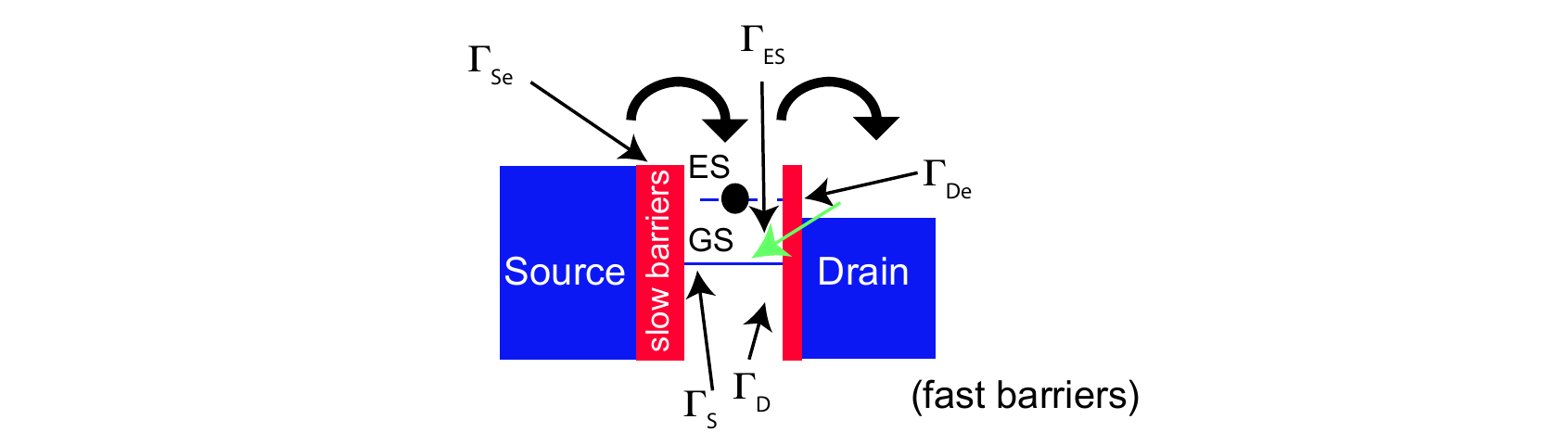}
\caption{Schematic of the processes that allow the observation of the excited state spectroscopy signature  of current in Fig.~5~b) of the paper. In green the previously neglected relaxation process where an electron tunnels from drain to the ground state, is shown.}
\label{fg:Fig}
\end{center}
\end{figure*}

\end{document}